\documentclass[sigconf,screen=true,anonymous=false,timestamp=true,acmthm=true,authordraft=false,authorversion=false,review=false]{acmart}

\usepackage{booktabs} 

\usepackage{xargs}                      
\usepackage{xcolor}  
\usepackage[prependcaption,textsize=tiny,textwidth=\marginparwidth]{todonotes}
\newcommandx{\unsure}[2][1=]{\todo[linecolor=red,backgroundcolor=red!25,bordercolor=red,#1]{#2}}
\newcommandx{\change}[2][1=]{\todo[linecolor=blue,backgroundcolor=blue!25,bordercolor=blue,#1]{#2}}
\newcommandx{\info}[2][1=]{\todo[linecolor=OliveGreen,backgroundcolor=OliveGreen!25,bordercolor=OliveGreen,#1]{#2}}
\newcommandx{\improvement}[2][1=]{\todo[linecolor=Plum,backgroundcolor=Plum!25,bordercolor=Plum,#1]{#2}}
\newcommandx{\makeinvisible}[2][1=]{\todo[disable,#1]{#2}}

\usepackage{threeparttable}

\copyrightyear{2018} 
\acmYear{2018} 
\setcopyright{rightsretained} 
\acmConference[PEARC '18]{Practice and Experience in Advanced Research Computing}{July 22--26, 2018}{Pittsburgh, PA, USA}
\acmBooktitle{PEARC '18: Practice and Experience in Advanced Research Computing, July 22--26, 2018, Pittsburgh, PA, USA}
\acmDOI{10.1145/3219104.3219165}
\acmISBN{978-1-4503-6446-1/18/07}

\begin{document}
\title{Leveraging OpenStack and Ceph for a Controlled-Access Data Cloud}

\author{Evan F. Bollig}
\affiliation{%
  \institution{MN Supercomputing Institute\\University of Minnesota}
  \streetaddress{599 Walter Library\\110 Pleasant St SE }
  \city{Minneapolis} 
  \state{Minnesota} 
  \postcode{55455}
}
\email{bollig@umn.edu}

\author{Graham T. Allan}
\affiliation{%
  \institution{MN Supercomputing Institute}  }
\email{gta@umn.edu}

\author{Benjamin J. Lynch}
\affiliation{%
  \institution{MN Supercomputing Institute}  }
\email{blynch@umn.edu}

\author{Yectli A. Huerta}
\affiliation{%
  \institution{MN Supercomputing Institute}  }
\email{yhuerta@umn.edu}

\author{Mathew Mix}
\affiliation{%
  \institution{MN Supercomputing Institute}  }
\email{mattmix@umn.edu}

\author{Edward A. Munsell}
\affiliation{%
  \institution{MN Supercomputing Institute }  }
\email{emunsell@umn.edu}

\author{Raychel M. Benson}
\affiliation{%
  \institution{MN Supercomputing Institute }  }
\email{bens0352@umn.edu}

\author{Brent Swartz}
\affiliation{%
  \institution{MN Supercomputing Institute }  }
\email{swartzbr@umn.edu}

\renewcommand{\shortauthors}{E. F. Bollig et al.}
\newcommand{\note}{\footnote}

\begin{abstract}

While traditional HPC has and continues to satisfy most workflows, a new generation of researchers has emerged looking for sophisticated, scalable, on-demand, and self-service control of compute infrastructure in a cloud-like environment. Many also seek safe harbors to operate on or store sensitive and/or controlled-access data in a high capacity environment.

To cater to these modern users, the Minnesota Supercomputing Institute designed and deployed Stratus, a locally-hosted cloud environment powered by the OpenStack platform, and backed by Ceph storage. The subscription-based service complements existing HPC systems by satisfying the following unmet needs of our users: a) on-demand availability of compute resources; b) long-running jobs (i.e., $> 30$ days); c) container-based computing with Docker; and d) adequate security controls to comply with controlled-access data requirements.

This document provides an in-depth look at the design of Stratus with respect to security and compliance with the NIH's controlled-access data policy. Emphasis is placed on lessons learned while integrating OpenStack and Ceph features into a so-called ``walled garden'', and how those technologies influenced the security design. Many features of Stratus, including tiered secure storage with the introduction of a controlled-access data ``cache'', fault-tolerant live-migrations, and fully integrated two-factor authentication, depend on recent OpenStack and Ceph features.
\end{abstract}

%
%





\maketitle

\section{Introduction}


Cloud Computing has taken the world by storm. For most, the Cloud promises immediate, on-demand access to resources, self-service management of infrastructure, plus the ability to operate with a level of agility and isolation not possible on traditional high performance computing (HPC) systems. Institutions like the Minnesota Supercomputing Institute (MSI)---primary provider for research computing services at the University of Minnesota---are fully aware of the allure of Cloud Computing to many of their users. 

For many years, MSI has sought to provide a few fully-managed HPC clusters as a shared resource for users, unified by a single global file system, and centralized user authentication. At present, MSI hosts two clusters, the latest of which, Mesabi, was purchased in 2015 and is still among the top 20 university-owned supercomputers in the nation. These clusters are available to all MSI users as batch scheduled resources, with the scheduler tracking a number of fair-share parameters to decide how and when jobs are run. Although Mesabi is predominantly homogenous in architecture and networking, heterogeneity does exist to meet diverse user needs in the specific configurations of memory (e.g., nodes range between 64 GB and 1 TB), as well as the presence of solid-state drives and/or accelerator boards. 

Whereas traditional HPC continues to satisfy the majority of needs across research disciplines, there is a growing trend toward data-intensive work with ever larger storage and compute requirements, as well as more stringent data-use agreements. Data-use agreements in particular are cumbersome to satisfy on HPC clusters where one-off modifications necessary to satisfy agency requirements for logging, backups (or lack thereof), and isolation would invasively impact the workflows and user-experiences of all users in the shared resource. The overhead in additional maintenance, monitoring, and user education may be acceptable for a small number of edge-cases, but the burden snowballs quickly to become unsustainable. 

Compounding the issue, the popularity of Cloud Computing has given rise to more sophisticated researchers who cross the role of developer with operations. These users, referred to as DevOps, seek an environment unchained from managed infrastructure, where they have the rights to escalate privileges to install software, manage firewalls, and control other aspects of system configuration, all in the name of agility and productivity. 


Facing the need to accommodate and sustain this trend, 
MSI designed and built Stratus, an on-premise cloud for protected data. Stratus is a subscription-based Infrastructure-as-a-Service that enables users to operate within their own virtual machines (VMs). Stratus is powered by the Newton version of the OpenStack \cite{openstack:2017} cloud platform, and is backed by the Luminous release of Ceph \cite{ceph:2006} storage. Stratus was opened to MSI users at the start of fiscal year 2018 (July, 2017). 
Beyond managing protected data, Stratus offers three 
features not present in MSI's existing cyberinfrastructure: a) on-demand availability of compute resources with self-service user control; b) long-running jobs (e.g., $ > 30$ days) that are live-migrated between compute nodes, and persist through maintenance periods; and c) container-based computing with Docker, which comes pre-loaded on MSI-blessed VMs. 

OpenStack was chosen for multiple reasons: a) it has built-in support for multi-tenancy, software defined networking, logging, SSL/TLS encrypted traffic and other features to make compliance easier; b) the modular OpenStack services enable mix-and-match configuration and are decoupled for fault tolerance; and c) a massive open source community 
distinguishes the project as the leading software for cloud deployments across industry and academia. Ceph is free, open source, and backed by much of the same community as OpenStack. Ceph excels as an efficient, low-cost storage platform with the ability to scale to many PBs. The stability, scalability, and value of OpenStack and Ceph have been vetted by large deployments like CERN \cite{Cern:2015} and the NSF funded JetStream \cite{Jetstream:2015}.  

In its initial release, Stratus is designed expressly to satisfy the NIH Genomic Data Sharing (GDS) Policy for the Database of Genotypes and Phenotypes (dbGaP) \cite{dbgap:2007}. Although other more stringent regulations and standards exist (e.g., HIPAA, FISMA, and ITAR to name a few), tackling dbGaP data 
is a good launching point for Stratus in that it addresses basic needs for access logging, data encryption, two-factor authentication, etc. Furthermore, it satisfies an immediate need at the University of Minnesota where researchers have no alternative available to analyze dbGaP data. Compliance with other regulations can be added in the future.


Stratus is most similar in spirit to the Bionimbus Protected Data Cloud (PDC) \cite{Bionimbus:2014}; a fully GDS- and HIPAA-compliant platform at the University of Chicago. Both Stratus and Bionimbus PDC are built on OpenStack and have Ceph storage. Stratus storage is entirely Ceph, in contrast to the current Bionimbus PDC, which has two object stores for protected data: 400 TB Ceph S3 and 1.7 PB IBM CleverSafe S3. Furthermore, Bionimbus PDC is a NIH Trusted Partner and allowed to maintain a complete persistent clone of the dbGaP data, while Stratus presents users a multi-tiered storage environment with a ``dbGaP Cache'' to store only active subsets of the dbGaP data in a scratch-like space. When the object lifecycle is complete or data becomes stale, they are deleted from the object store. 
This allows Stratus to avoid bit rot, and conserve storage costs. 

In a similar vein, the CancerCollaboratory \cite{CancerCollab:2017} at the Ontario Institute for Cancer Research, is a Ceph-backed OpenStack cloud for cancer-related data. CancerCollaboratory is designed for maximum storage capacity, which is used to persistently clone, among other things, dbGaP data dealing with cancer from Bionimbus PDC. 

The NIH has also sponsored three public cloud pilots: a) Broad Institute FireCloud \cite{FireCloud:2017}; b) Seven Bridges Cancer Genomics Cloud \cite{SevenBridges:2017}; and c) the Institute for Systems Biology Cancer Genomics Cloud \cite{ISBCGC:2017}. All three pilots run on public providers (Google/Amazon), where the annual cost of computing is substantially higher than on a local on-premise cloud. To alleviate this burden, the NIH is currently offering credits for approved projects to spend at the public providers.

\section{Stratus Cloud Platform}

Stratus is a relatively small cluster, less than one rack in size. 
The Stratus compute hardware is currently twenty HPE ProLiant XL230a nodes, each with two Intel E5-2680v4 (14-core) CPUs, 256 GB of RAM, and 10 GbE networking to the outside world. Hyper-threading is enabled for a total core count of 1120 cores. An array of eight HPE Apollo 4200 storage servers are each connected to compute nodes via two redundant 40 GbE switches. Each storage node has 256 GB memory, two 800 GB Intel NVMe P3700 accelerator boards, 
and 198 TB raw capacity; yielding 1.5 PB raw capacity for the entire cluster. Ten additional 8-core Advanced Clustering control servers, each with 32 GB RAM, run the OpenStack APIs, monitor storage operations and orchestrate VM lifecycles. Two of the ten are network controllers for VM traffic and have dual 40 GbE network connections, while the remaining eight nodes have dual Intel X520 10GbE network, and various purposes: two to run the Ceph RADOS gateway, three Ceph storage monitors, one primary OpenStack API and identity controller, one OpenStack storage controller, and one telemetry and admin node.

The Stratus platform is comprised of OpenStack services: Keystone, Cinder, Glance, Nova, Neutron, and Horizon. Keystone manages projects, groups and user role assignments. Identity management is outsourced by Keystone to the central UMN Shibboleth server, also requiring Duo two-factor authentication \cite{Duo:2017}. The Cinder API provides the ability to create both storage and VM boot volumes within Ceph block device storage. The Glance API publishes a repository of MSI-blessed machine images to all projects, though denies permissions to create/modify images for regular users. Since Glance images reside in block device storage, users can create boot volumes as Copy-on-Write clones of these images, and boot VMs in seconds. Cinder snapshots of boot volumes suffice as machine images in the user-space and count toward project volume quotas. Nova manages virtual machines on compute nodes. VMs are run with QEMU+KVM hardware accelerated virtualization for minimal overhead. Although CPU pinning is a tuning option to optimize compute-intensive workloads, it is incompatible with live-migration, which is critical for Stratus to support long running jobs. Nova is configured to oversubscribe hardware by a factor of 2x for vCPUs and 1x for memory. Neutron BGP dynamic routing allows VMs to be assigned to campus routable RFC1918 IP subnets and have their gateways automatically discovered by the network. They can then communicate directly to other campus resources without any Network Address Translation (NAT). Access to external resources is provided by an upstream NAT device for outbound routes, but no ingress connections are allowed. In a future version of OpenStack, this configuration will allow floating IP addresses to bypass the network nodes to send packets directly to the hypervisor host where the VM is currently located. The Horizon web interface, which requires two-factor authentication to enter, binds these separate web services into a single simple and intuitive portal where users can perform self-service management of their infrastructure.
For access to the OpenStack command line interface or to connect to VMs, users must first SSH to the Stratus Bastion server and authenticate with two factors.

\section{Secure VMs with MSI-Blessed Images}

\begin{figure}[!t]
\centering
\includegraphics[width=2.85in]{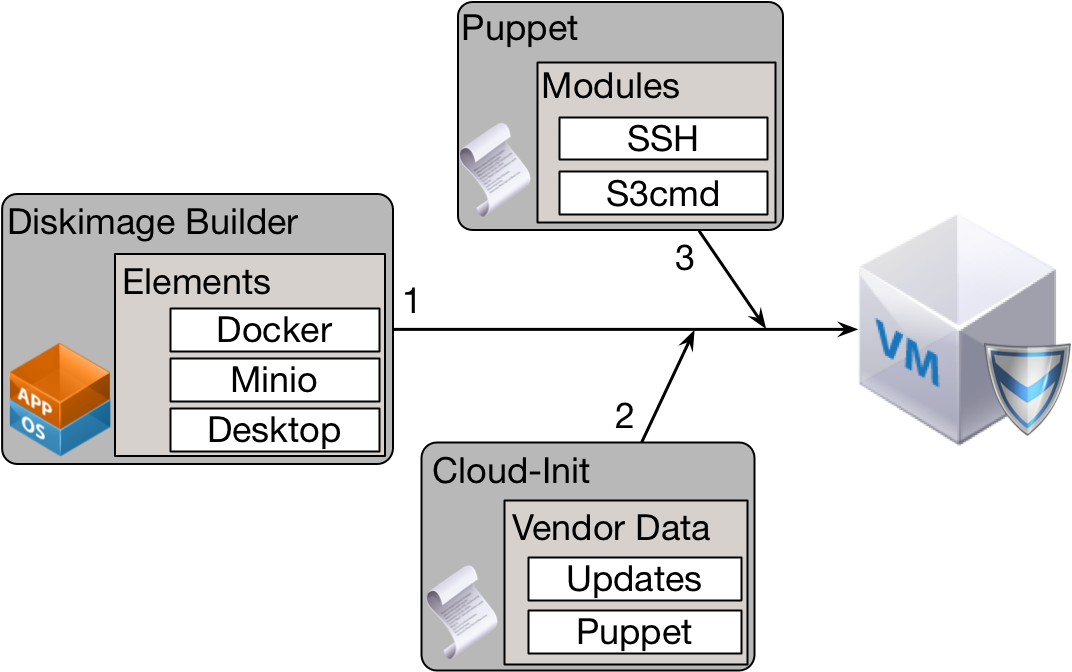}
\caption{Secure VMs start with ``MSI-blessed'' images constructed by DiskImage Builder. Cloud-Init and Puppet ensure that VMs are updated with the latest security patches and configured for compliance at boot.}
\label{fig:dib_images}
\end{figure}

MSI provides a number of customized base images for users to boot when creating a VM on Stratus. All images are Linux distributions such as Debian and CentOS. The primary benefit from users leveraging MSI-provided customized images, versus the vanilla images direct from Linux distributions, is that MSI can enable basic security controls for users and have peace of mind that a VM is protected until users opt-out (at their own risk). Generating images also presents an opportunity to properly configure remote desktop (XRDP), and other heavy-weight research tools such as Galaxy \cite{afgan2016galaxy}, which is a web-based workflow manager popular in the genomics and proteomics communities. 

There are three components that contribute to the images that users interact with as illustrated by Figure~\ref{fig:dib_images}. First the image is built with DiskImage Builder \cite{dib:2018}, a tool created and supported by the OpenStack community. DiskImage Builder composes images from a prescribed set of features called Elements which each describe how to provision individual softwares, configuration files, accounts, etc. DiskImage Builder is extensible and MSI has generated a collection of custom Elements to setup tools like the Minio Client and ensure that Remote Desktop, and Docker run at boot. 

When a VM boots, a service runs at startup called Cloud-Init \cite{cloudinit:2018}. Cloud-Init pulls metadata from the OpenStack Nova metadata server and configures settings at first boot like IP address, user keys, etc. Cloud-init also has flexible mechanisms (e.g., vendor data) for MSI to inject instructions and settings. In this case, Cloud-Init forces VMs booted from MSI-blessed images to apply system security updates. Security updates are the first field of defense in ensuring the integrity of VMs and protecting all users from potential compromise.  

The final component contributing to secure a VM based on MSI-blessed images is Puppet \cite{puppet:2018}. Puppet is provisioned in the image by a DiskImage Builder element and runs as a daemon service on the instance after boot. It periodically checks a master server for reference configurations and will apply changes to the virtual machine to match the reference. This ensures a base state on the virtual machine such as user- and group-permissions, and ssh keys. Although Puppet is traditionally used to fully manage infrastructure, this use-case is lightweight and intended for emergency patching when necessary. 

A strict naming convention is applied to images so users know what Elements go into each image and the Linux distribution it is based on. For instance, ``Centos7\_dbgap\_blessed\_desktop'' describes an image based on the CentOS 7 distribution, with dbGaP related tools (e.g., gdc-client and aspera) for pulling dbGaP data from the NIH, and Remote Desktop (``desktop'') running at boot.

MSI currently maintains 9 images. The available Linux distributions are CentOS 7, Debian 9, and Ubuntu 16.04, and each distribution has Vanilla (i.e., the original image provided by the the distribution) and dbGaP-blessed (MSI) variants. For the CentOS 7 distribution, MSI also offers variants to integrate Remote Desktop and Galaxy services.

New images are periodically released as major security issues are patched by distributions, or new features are requested by users. The image build-process is automated, and as new images are uploaded, old images are phased out. 

%

\section{Stratus Storage}

\begin{figure}[!t]
\centering
\includegraphics[width=\columnwidth]{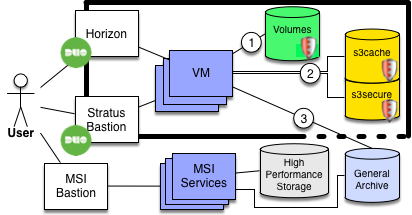}
\caption{Stratus is sandboxed from other MSI services, but VMs have access to three tiers of storage. Shields denote suitable for dbGaP data.}
\label{fig:stratus_sandbox}
\end{figure}

MSI selected the HPE Apollo 4200 server as the storage building block for Stratus due to its high storage density, without being too large to assemble a reasonable number of nodes for a viable Ceph cluster. Each node supports 24 8 TB hard drives, six 960 GB SSDs and two 800 GB NVMe flash cards in 2U of rack space. Each individual hard drive or SSD is the basis of a single Ceph object storage daemon (OSD). The NVMe flash provides high speed journal space for the hard drive-based OSDs which provide the majority of the storage, reducing their write latency. Each NVMe flash card supports journaling for 12 hard drive OSDs. The smaller set of SSD-based OSDs contain indexes for the s3 object stores, as well as an optional high performance block storage pool.

The drives are configured on the server as individual single-drive RAID-0 units, rather than in ``JBOD'' pass-through mode, in order to leverage the onboard cache of the RAID controllers. While the servers RAID controller supports hardware encryption of the connected storage, it was decided to avoid this in favor of Ceph's existing support for linux standard LUKS encryption. This was motivated by several reasons: a) this kept the setup as hardware-agnostic as possible; b) testing showed a negligible performance impact (roughly ~1\%); and c) it was felt that Ceph's encryption key management was superior, storing the keys away from the storage node in the Ceph monitor key-value store.

Within Ceph, Stratus storage is partitioned into three areas: a) a 3x replicated block device storage for Cinder volumes with 200 TB usable capacity; b) an S3 object store for caching dbGaP data with 500 TB capacity; and c) a persistent S3 object store for secure archive. Both b) and c) are 4:2 erasure-coded. Each of these storage classes uses its own set of underlying ceph storage pools, on which quotas are enforced to maintain the desired partitioning.

S3 object lifecycle (expiration) was added as a feature in Ceph Kraken, and is essential to the design of the dbGaP Cache. Lifecycle policies are applied to buckets and impact all objects within. Although this is still experimental, the plan is to enforce a default expiration of 60-days, such that objects are deleted following a First-In-First-Out policy. If greater than 80\% of the dbGaP Cache is filled, we gradually reduce this threshold until utilization falls below the 80\% mark. While a Least Recently Used lifecycle policy might be seen as preferable, s3 storage semantics make this unlikely to be possible; object creation being the only stored timestamp.

From the user perspective Stratus has three storage tiers, which are illustrated in Figure~\ref{fig:stratus_sandbox}. Volumes are the first tier, and the layer where active data processing occurs. The second tier, secure archive, is the combination of Stratus' two secure object stores where users can target either dbGaP Cache (``s3cache'') or persistent secure storage (``s3secure'').  Users interact with this tier using an s3 client; our recommendation being Minio Client (``mc''), both for ease of use, and because recent versions support server-side data encryption using customer-provided keys (SSE-C). A third ``general archive'' tier, is MSI's general-purpose S3 object store on a separate dedicated Ceph cluster . It is through this general archive that data can be brought in from other MSI resources. Stratus users can move data here when it is sanitized and would not violate the data-use agreement to be stored in an unprotected environment.

A typical workflow begins with using the Horizon web interface to boot a VM with a small boot volume plus a large (e.g. 1 TB) workspace volume. Next, the user connects to the VM via the Stratus Bastion server and transfers the desired dbGaP data set from the NIH endpoints to the workspace volume for analysis. The dbGaP data can optionally be copied into the s3 dbGaP cache for later reuse without retransfer from NIH. Intermediate data produced during analysis can also be stashed in the dbGaP Cache, while final results can be stored in persistent secure storage. At any point during analysis, the user can grow the size of the workspace, snapshot it, or detach it from one VM and attach it elsewhere (e.g., if workflow stages are setup as separate VMs). When analysis is complete, the user can post sanitized data to MSI's general-purpose S3 storage where they can access it from other MSI resources or share with colleagues.

A similar workflow is adopted by users of the Galaxy machine image on Stratus. In those cases, the Galaxy software stores data in the dbGaP Cache, and performs read and write operations on a local volume with small quota (~200 GB up to 2 TB). As the quota is reached, data on the local volume is purged automatically by Galaxy. The benefit of dbGaP Cache in this case is that users can process significantly more data with a small VM footprint. 

In future, several improvements to the storage subsystem may be investigated. For example, Ceph's new ``bluestore'' storage backend is now considered stable, and offers potential performance improvements. The ``Queens'' release of OpenStack adds native LUKS encryption of volumes, which would permit compliance with more stringent data-use agreements by supporting customer-specific encryption keys on active volumes. Another feature anticipated in Queens is multi-attach volumes, which would enable users to attach their large workspace volume to more than one VM simultaneously, and vastly simplify multi-VM workflows. The presence of multi-attach could dramatically impact the level of block storage utilization.

%






\section{HPC-like Performance}

Once the Stratus production system was deployed, MSI invested a significant amount of time to benchmark the system and assess the prospect of achieving HPC-like performance on the cloud. As a baseline for HPC performance expectations, the performance on Stratus matched against the performance on Mesabi, MSI's flagship HPC. 

\begin{figure}[!t]
\centering
\includegraphics[width=\columnwidth]{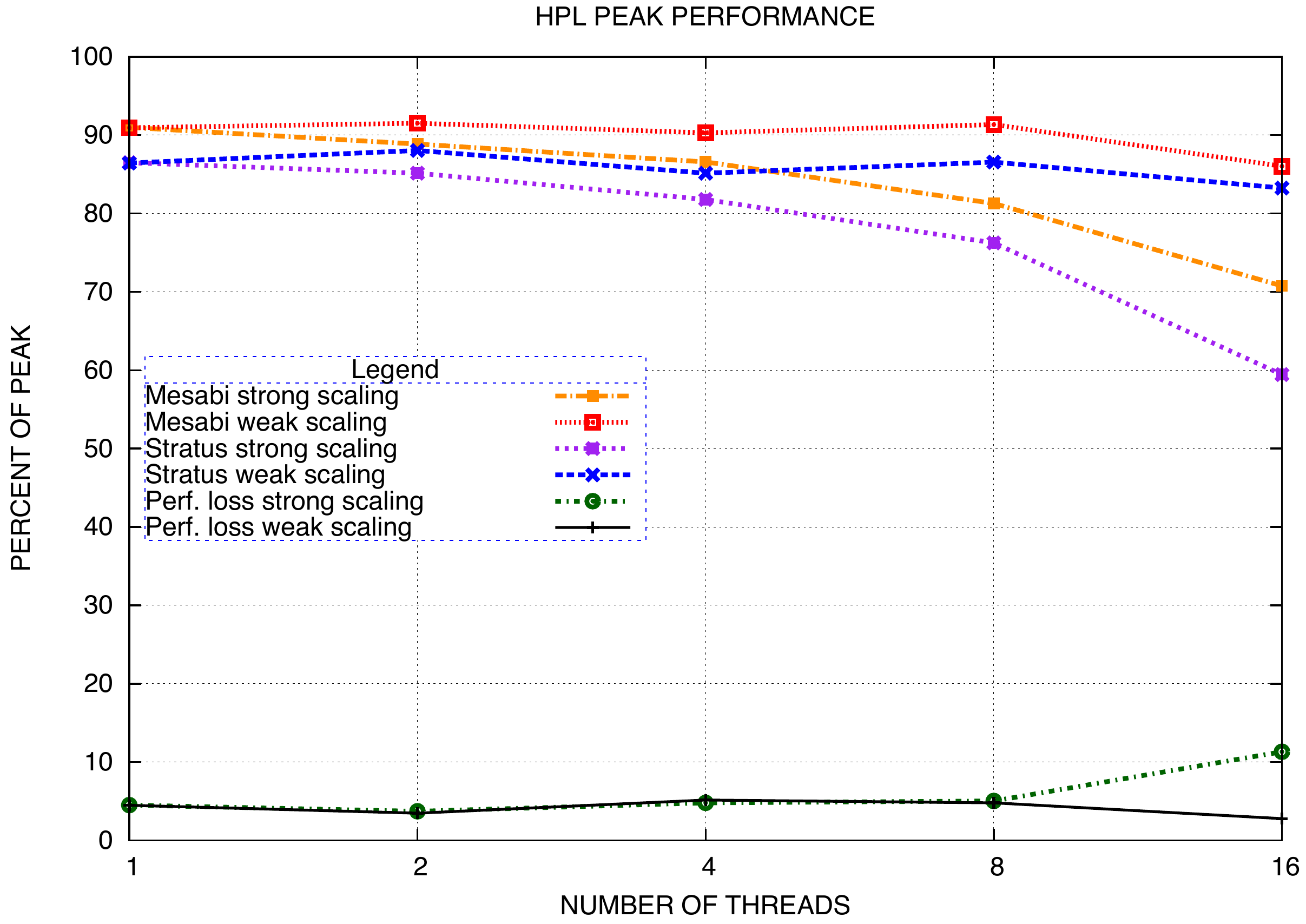}
\caption{HPL peak performance on a bare metal HPC-node of Mesabi versus a cloud-based VM running on Stratus.}
\label{fig:hpl_performance}
\end{figure}
Figure~\ref{fig:hpl_performance} compares the HPL \cite{hpl2018} peak performance on a bare metal node (Mesabi) and a 16 vCPU VM (Stratus) running on a 28 core Broadwell node). Weak scaling makes the problem size per thread constant, so that the amount of work per thread stays constant as the number of threads is increased. Whereas strong scaling fixes the problem size, and that work is spread over increasing numbers of threads, so the amount of work per thread decreases as the number of threads is increased. Weak scaling used a matrix size that used 80\% of 2 GB/thread, which is calculated as $N=\sqrt{(t * 0.8 * 2 * 1024^3)/8}$ where $t$ is the number of threads specified for the benchmark. Note that the root is required because N represents one side of a square matrix. The strong scaling runs used an N based on 80\% of 2 GB. The choice of 2 GB/thread was arbitrary, but was chosen to fill most of the available memory on Mesabi's 64 GB nodes when the number of threads is scaled up to 24, and also ensures the HPL calculation is not entirely resident in cache.

The yellow and red curves show the strong and weak scaling peak performance attained using HPL on a bare metal Haswell (dual 12 core 2.5 GHz Xeon E5-2680 v3) node (Mesabi), which has a peak performance of 40.0 GFLOPS/core. At 1 thread, peak performance of 91\% shows very reasonable peak performance. Within the HPC community, getting over 90\% of peak for HPL performance is generally considered excellent performance. 
The purple and blue curves show the strong and weak scaling peak performance attained using HPL on a cloud based Broadwell (dual 14 core 2.4 GHz Xeon E5-2680 v4) 16 vCPU VM (Stratus), which has a peak performance of 46.4 GFLOPS/core. At 1 thread, peak performance of 87\% shows an approximate 5\% loss due to virtualization overhead relative to the bare metal Haswell node. It is worth noting that for the Mesabi system, lscpu reports an L3 cache of 30 MB for the system, but within the virtualized system no L3 cache was reported. This missing tier of cache could certainly degrade the performance by affecting the VM kernel scheduler, resulting in the scheduler generating unnecessary inter-processor interrupts.


For the red weak scaling curve, the peak performance is maintained at about 91\% as we increase the number of threads, indicating good scaling behaviour, until 16 threads is reached, when performance reaches 86\%. This reduction to 86\% indicates that even the weak scaling case does not scale all the way up to 16 threads, likely due to memory/cache contention on the node. Likewise, the blue weak scaling Stratus curve maintains good scaling behaviour at about 87\% peak performance out to 8 threads, then dips to 83\% at 16 threads. The performance loss for the weak scaling runs is shown by the black curve, i.e. the Mesabi bare metal weak scaling peak performance minus the Stratus VM weak scaling peak performance. It shows that weak scaling peak performance is fairly consistent across the number of threads, at around 5\% loss due to virtualization.

For the yellow strong scaling curve, performance decreases as the number of threads is increased, which is normal strong scaling behaviour. The strong scaling behaviour for the Stratus VM shown by the purple curve is similar to the bare metal Haswell node, except when the number of threads is 16, the strong scaling peak performance reaches 59.4\%, mentioned below. The performance loss for the strong scaling runs is shown by the green curve, the strong scaling analogue to the black curve. The performance loss for the strong and weak scaling are consistent at about 5\%, until 16 threads are reached, when the strong scaling performance loss increases to 12\%, while the weak scaling performance loss decreases to 3\%. At 16 threads, the strong scaling on the virtualized resource is performing worse. However, this is not of much concern since the shrinking problem size is already unable to saturate the core, so any amount of noise in the solution is going to heavily impact the performance. In fact, we see a significant increase in the standard deviation as we increase the number of threads from 8 to 16.

\begin{figure}[!t]
\centering
\includegraphics[width=\columnwidth]{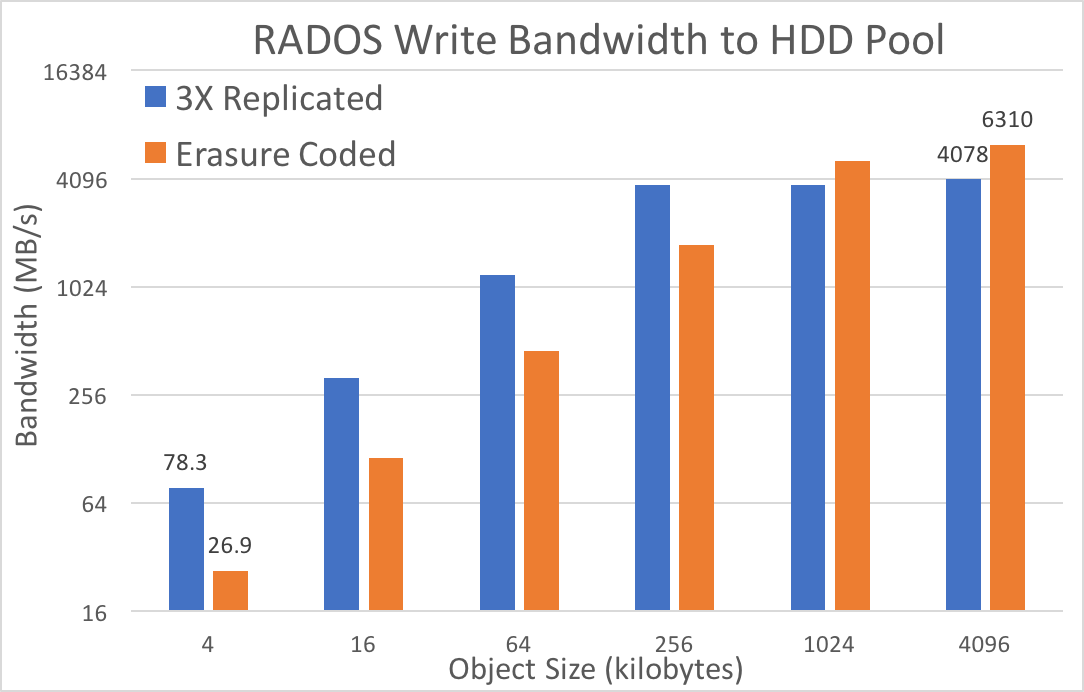}
\caption{Stratus storage write bandwidth for both 3x replicated block device storage and erasure coded object storage.}
\label{fig:storage_bandwidth}
\end{figure}
Figure~\ref{fig:storage_bandwidth} compares the performance achieved by the 3x replicated pool to the erasure-coded pools on the cluster. For this test, ten compute nodes were each running the RADOS benchmark tool with eight concurrent operations. The 4:2 ($k:m$) configuration of the erasure-coded pool 
uses 1.5 MB of space in backend storage for every MB of data written to the pool. The 3x replicated pool writes 3 MB of data to the HDD for every MB written. 
The replicated pool outperforms the erasure-coded pool for small writes. In addition to the added CPU load from object storage, every object stored on the erasure-coded pool writes a total of 6 objects ($k+m$) to the HDD-backed OSDs. This leads to the observed performance in the IOPs-limited regime. For large objects ($>1$ MB), the erasure-coded pool finally outperforms the replicated pool. At the 4 MB object size, the 3x replicated pool is writing 12,234 MB/s to the backend storage ($4078 * 3$), and the erasure-coded pool is only writing at 9,465 MB/s ($6310 * 1.5$). 

In terms of observed performance within VMs, a number of FIO \cite{fio2018} storage benchmarks on Stratus showed the Volume storage capable of achieving 270 MB/s in write bandwidth. This performance is impressively 12.5\% higher than the 240 MB/s write bandwidth achieved by FIO on Mesabi's high performance filesystem. Obviously, there are major scaling differences between the two systems, and they have been architected for different purposes. However, Stratus certainly demonstrates HPC-like performance in storage as well as compute.

\section{Stratus Subscription}

Due to the specialized nature of Stratus, the small size, and the potential need for growth, it was decided that it should be offered as a paid service. An annual subscription fee was developed using Cornell's Red Cloud \cite{RedCloud:2017} as a model. The base subscription costs less than a thousand dollars per year, but allots users 16 vCPUs, 32 GB RAM, 2 TB of Volume storage, and access to the shared dbGaP Cache for short-term data storage. Users are also granted access to a number of MSI-blessed CentOS and Ubuntu base images, and sudo permissions to install software on personal virtual machines. 
Additional vCPUs, Volume storage, and persistent object storage can be added \`{a} la carte in 1 vCPU or 1 TB/year increments over the base subscription. Subscription pricing is calculated based on zero-profit cost recovery of the hardware, including all network switches, and all control-, storage-, and compute-nodes. Staff time for operational/administrative tasks and user support are also factored into the price. Network traffic is excluded. Note that the subscription ensures cost-recovery of the system at 85\% utilization and allows for continuous expansion as popularity grows. The presence of a fee also ensures that most (not all) users will exercise good cloud hygiene and actively clean-up/scale-in their resources when not needed.


This new model has three benefits: a) as the popularity of Stratus grows, cost recovery ensures funds will be available to scale the cluster; b) it encourages casual users to continue working on traditional HPC systems where MSI provides cycles and limited storage for free; and c) it encourages serious users to practice good cloud hygiene by securing the space they are paying for, and to clean up after themselves. Although the base subscription is generous, all VM boot volumes, data volumes and snapshots count against the 2 TB Volume allocation, and that space can fill quickly. 

Subscriptions are mapped into OpenStack using built-in service quotas. OpenStack attaches quotas to individual projects, with toggles for vCPU count, size of volumes and snapshots, RAM, etc. Object storage quotas are applied at the bucket-level on Ceph.

%
%


\section{GDS Compliance}

Most cloud providers absolve themselves from responsibility of protecting users, focusing instead on the minimum effort necessary for regulatory compliance within the underlying hardware, and leaving users to individually manage and vet whatever is run on top of the cloud. This makes sense in a self-service Infrastructure-as-a-Service model, but creates a potentially dangerous situation for novice users with protected data. Stratus was built with an MSI-first mentality, prioritizing security of the hardware, OpenStack, and Ceph. However, as an academic unit supporting computationally intensive research at the University of Minnesota since 1984, MSI has a culture of going the extra mile to train, protect, and assist. In this case, a fair amount of effort also went into sandboxing user VMs and providing sane defaults for security that users can opt-out of at their own risk. 
This included following the best practices guide for GDS data \cite{dbgapBestPractices:2015} as a checklist. All bullets therein were treated as required controls. Note that the NIH best practices go beyond what most large research computing facilities can or need to do.

\begin{figure}[!t]
\centering
\includegraphics[width=\columnwidth]{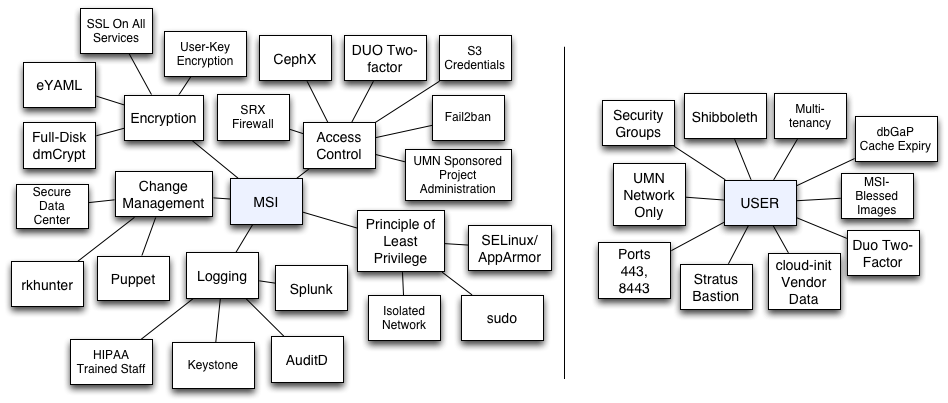}
\caption{Compliance is met with an MSI-first policy that focuses on securing MSI, OpenStack, and Ceph. Users are protected with sane defaults and network security.}
\label{fig:msi_first}
\end{figure}
Figure~\ref{fig:msi_first} illustrates a sampling of protections for MSI and users as two trees. While outside the scope of this document to exhaustively detail each measure, there is clearly a bias toward protecting MSI. The MSI tree on the left has two levels of nodes. The first level are five significant controls mentioned by the best practices guide (i.e., Encryption, Logging, Principle of Least Privilege, Access Control, and Change Management), and the second level summarizes some of the ways we have addressed them. The majority of the GDS controls are intangible to users, but represent substantial administrative and operational staff effort. 

\begin{table}[!t]
\centering
  \caption{Firewall restrictions for Stratus}
  \label{tab:firewall_restrictions}
  \begin{threeparttable}
    \begin{tabular}{ | l | l | l |}
    \hline
    Scope & Ingress & Egress  \\ \hline\hline
    World & None & ALL \\ \hline
    UMN & 443, 8443 with SSL\tnote{1} & ALL \\ \hline
    Stratus & ALL via Stratus Bastion\tnote{1} & ALL \\ \hline
    Project & ALL & ALL \\ \hline
    \end{tabular}
    \begin{tablenotes}
    \item[1] requires Security Group exceptions
    \end{tablenotes}
   \end{threeparttable}
\end{table}
The USER tree offers measures that users are aware of. A few relate to GDS controls (e.g., two-factor authentication), but others go above and beyond. The most impactful measures deal with network traffic. In most clouds, users get unrestricted traffic in and out of VMs. 
Table~\ref{tab:firewall_restrictions} presents firewall rules---both ingress and egress---for VMs on Stratus. In all cases, egress traffic is unrestricted, but ingress is only open between VMs within the same project, or from the Stratus Bastion (so long as Security Group exceptions exist). Traffic is blocked on all ports from the world, but when it originates from the campus network, ports 443 and 8443 are accessible with caveats: a) SSL is required on both, and b) users must create a Security Group exception to opt-out of the default network protection. These rules sandbox Stratus VMs away from other MSI services, the University, and the world. For VMs requiring world-wide access, MSI recommends the NIH cloud pilots \cite{FireCloud:2017,ISBCGC:2017,SevenBridges:2017}. 


Encryption is essential for data obfuscation and protection, and it is leveraged on many levels including full-disk encryption of storage disks, encrypted network traffic, and user-encrypted data at rest. Access controls limit who can interact with systems, software, and data, and also establish how users are properly identified (e.g., with two-factors). Principle of Least Privilege prevents administrators and users from running software, accessing data, or completing any other tasks for which they are not approved. Logging is mandatory and encompasses tracking account access, data access, software installations, account escalations, violations, etc. Logging also extends to parsing, management, and protection of log integrity. Lastly, Change Management tracks modifications to Stratus including web service configurations, software updates, as well as physical hardware changes.

\section{Conclusion}
Technical details of Stratus, a subscription-based Infrastructure as a Service for research computing, were presented herein. Stratus satisfies the immediate need for an environment to securely process dbGaP data at the University of Minnesota, and meets or exceeds the requirements laid out by the NIH Genomic Data Sharing Policy.

Future work will consider compliance with other regulations, as well as supporting use cases with unrestricted data types. This may include cluster solutions for processing dbGaP data (e.g., projects like Kubernetes). Also, MSI is currently piloting a general-use partition within Stratus for non-dbGaP projects and open-access data types. Projects under the general space are separated from dbGaP users by a second neutron provider, and subject to less stringent access controls and firewall rules. 




\bibliographystyle{ACM-Reference-Format}
\bibliography{bibliography}


\end{document}